\documentclass[10pt]{article}

\usepackage{pdfpages}
\usepackage{float}
\usepackage{amsmath}
\usepackage{mathtools}
\usepackage{amsfonts}
\usepackage{jheppub}
\usepackage{color}
\relax
\newcommand\tree{\rm tree}
\newcommand{\beq}{\begin{equation}}
\newcommand{\eeq}{\end{equation}}
\newcommand{\beqn}{\begin{eqnarray}}
\newcommand{\eeqn}{\end{eqnarray}}
\newcommand{\slsh}{\rlap{$\;\!\!\not$}}     
\def\spa#1.#2{\left\langle#1\,#2\right\rangle}
\def\spb#1.#2{\left[#1\,#2\right]}
\def\sp{\slsh{p}}
\def\braket#1{\langle #1 \rangle}
\def\Trfourgfive#1.#2.#3.#4{{\rm tr}_5\{{#1}\,{#2}\,{#3}\,{#4}\}}
\def\Trsixgfive#1.#2.#3.#4.#5.#6{{\rm tr}_5\{{#1}\,{#2}\,{#3}\,{#4}\,{#5}\,{#6}\}}
\def\TrsixgL#1.#2.#3.#4.#5.#6{{\rm tr}_{-}\{{#1}\,{#2}\,{#3}\,{#4}\,{#5}\,{#6}\}}
\def\TrsixgR#1.#2.#3.#4.#5.#6{{\rm tr}_{+}\{{#1}\,{#2}\,{#3}\,{#4}\,{#5}\,{#6}\}}
\def\TrfourgL#1.#2.#3.#4{{\rm tr}_{-}\{{#1}\,{#2}\,{#3}\,{#4}\}}
\def\TrfourgR#1.#2.#3.#4{{\rm tr}_{+}\{{#1}\,{#2}\,{#3}\,{#4}\}}

\title{On Higgs boson plus gluon amplitudes at one loop}
\author{
    R. Keith Ellis \& Satyajit Seth\\
    IPPP, Durham
    \\
    E-mail: 
    {\tt keith.ellis@durham.ac.uk, satyajit.seth@durham.ac.uk}.}

\preprint{IPPP/18/73}

\abstract{
We present analytic results for one-loop Higgs boson + $n$-gluon
amplitudes for $n \leq 5$ in the full theory including all dependence
on the (top) quark mass.  In this paper we consider only the case
where the gluons all have the same helicity.  The amplitudes are
expressed in simple formula and display similar structure.  Their
limiting behaviour in small Higgs momentum and large top mass is
studied.}


\begin{document}

\maketitle

\section{Introduction}
It is evident that detailed study of the Higgs boson will be a primary
focus of the experiments performed at the CERN LHC for at least the next decade. 
Many calculations of Higgs boson production by gluon fusion are carried out in the Higgs
boson effective field theory, valid when the top mass is larger than
all other scales in the problem. This approach has the merit that
calculations performed in the effective theory are easier, since the
Born-level matrix element is a tree graph, rather than a one-loop
process.  However, with increasing statistics the LHC will be able to
probe a regime where the effective theory is no longer valid, yielding
valuable information about the intermediaries circulating in the loop
that couple to the Higgs boson. This is the case in Higgs boson + jet 
production when the transverse momentum of the Higgs boson or of the jets
is large compared to the top quark mass.

Next-to-leading order (NLO) QCD corrections to Higgs boson plus 1-jet production with full
top-quark mass dependence are already known~\cite{Lindert:2018iug,Jones:2018hbb}. These
calculations use the one-loop Higgs boson + 3 parton amplitude as the
Born-level cross section, and the one-loop Higgs boson + 4 parton
amplitude as a real radiation correction to the Born-level process. 
The two loop virtual corrections are calculated
using an expansion method~\cite{Lindert:2018iug} 
or sector decomposition~\cite{Jones:2018hbb}.
If one were to go further and calculate the NNLO QCD corrections to the
Higgs boson + 1 jet process, 
one of the ingredients would be the Higgs boson + 5 parton amplitudes.  

The Higgs + 2 jet process via gluon fusion has also been
calculated at leading order in the full theory~\cite{DelDuca:2001eu,DelDuca:2001fn}.  
This process constitutes a ``background'' to the Higgs + 2 jet process occurring
via Vector Boson fusion, which also comes accompanied by two jets.
The leading order Higgs + 3 jet process has been considered in ref.~\cite{Campanario:2013mga}.
Phenomenological analyses of Higgs + jets including full mass effects have been 
performed in refs.~\cite{Greiner:2015jha,Harlander:2012hf}. 

Despite this progress the literature does not contain detailed
analytic results for Higgs + $n$-parton amplitudes in the full theory
for $n \geq 4$. Techniques for the analytic calculations based largely on 
unitarity have been developed over a number of 
years~\cite{Bern:1994zx,Bern:1994cg,Bern:1996je,Bern:1995db,Britto:2004nc,Forde:2007mi,Badger:2008cm}.   
The purpose of the current paper is to provide
analytic results for the specific case of gluons all having
the same helicity. We have undertaken this work, in order to elucidate
patterns which exist for varying values of $n$. In addition, by calculating 
the all positive helicity processes, which are the simplest, we can 
assess the feasibility of obtaining simple analytic forms for all helicities.

From a numerical point of view, one loop calculations are a solved problem thanks 
to techniques~\cite{Ossola:2006us,Cascioli:2011va} that allow numerical calculation of
the coefficients of the needed loop integrals\footnote{For a review and a complete set of references, 
see ref.~\cite{Ellis:2011cr}.}.
The full numerical result is obtained by combining these numerical
results for the coefficients, 
with analytic results for the one-loop scalar integrals.  Analytic
expressions for scalar one-loop integrals are completely known both
for IR-finite~\cite{tHooft:1978jhc} and divergent\cite{Ellis:2007qk}
cases. The downside of this semi-numerical approach is that it can lead to instabilities in
corners of phase space. These instabilities can be solved by moving to
higher precision calculation, at the cost of increased computer
time. Analytic calculations on the other hand are less prone to these
instabilities.

\section{Preamble: Unitarity calculation of $H+2g$ amplitude}
The aim of this paper is to calculate one-loop results for Higgs boson 
+ gluon amplitudes. These amplitudes contain a quark of mass $m$ circulating 
in the loop, (dominantly the top quark) and the coupling of the Higgs boson
to the quark is given by $-i m/v$ where $v \approx 246$~GeV is the 
vacuum expectation value of the Higgs field.
The mass of the Higgs boson is denoted by  $M_h$.

We will calculate colour-ordered sub-amplitudes for the production of a Higgs boson and $n$ gluons defined as follows:
\begin{eqnarray}
        \label{exp}
        {\cal A}_n(\{p_i,h_i,c_i\})\,&=&\,i\frac{g_s^n}{16 \pi^2} \frac{1}{v}\sum_{\{1,2,\dots,n\}'}\;tr\,(t^{c_1}t^{c_2} \dots t^{c_n})
        A_n(1_g^{h_1},2^{h_2}_g,\ldots n^{h_n}_g;H)\, ,
\end{eqnarray}
where the sum with the {\it prime}, $\sum_{\{1,2,\dots,n\}'}$, is over all
$(n-1)!$ {\em non-cyclic}  permutations of $1,2,\dots,n$ and the $t$ matrices are the SU(3) matrices in the fundamental representation normalized such that,
\begin{equation}    \label{normalization}
        tr(t^a t^b)\;=\; \delta^{ab}.
\end{equation}
Because of Bose symmetry it will be sufficient to calculate one
permutation, and the other colour sub-amplitudes can be obtained by
exchange.

The unitarity method seeks to calculate this result by sewing together
tree-level colour-ordered sub-amplitudes.  For the tree graph process,
$qgg\ldots g\bar{q}$, these are defined as,
\begin{eqnarray}
        \label{colorordered}
        {\cal G}^n_{ab}(p_a,h_a,\{p_i,h_i,c_i\},p_b,h_b) &=&\,i g_s^n \sum_{\sigma \in S_n} (t^{c_{\sigma(1)}}t^{c_{\sigma(2)}}
 \dots t^{c_{\sigma(n)}})_{ab}
        G_n^{\tree} (a_q,\sigma(1),\ldots \sigma(n),b_{\bar{q}}) \, ,
\end{eqnarray}
where $S_n$ is the permutation group on $n$ elements, and $G_n^{\tree}$ are the tree-level partial amplitudes. In a similar way we can define the 
tree-level sub-amplitudes for the production of a Higgs boson and gluons from a massive fermion line,
\begin{eqnarray}
        \label{colororderedHiggs}
        {\cal H}^n_{ab}(p_a,h_a,\{p_i,h_i,c_i\},p_b,h_b)\,&=&-i \frac{g_s^n}{v} \sum_{\sigma \in S_n} (t^{c_{\sigma(1)}}t^{c_{\sigma(2)}}
 \dots t^{c_{\sigma(2)}})_{ab}
        H_n^{\rm tree} (a_q,\sigma(1),\ldots \sigma(n),H,b_{\bar{q}}) \, .
\end{eqnarray}

For the case of Higgs + 2 gluons the only non-zero amplitude is when 
the gluons have the same helicity. 
\begin{figure}[t]
\begin{center}
\includegraphics[width=4cm,angle=270]{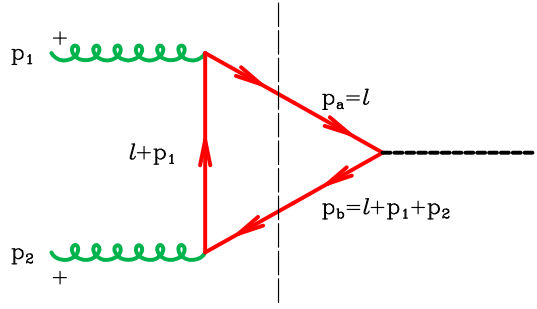} \\
\caption{\label{Hgg} Unitarity approach to calculating the Higgs + 2 gluon amplitude}
\end{center}
\end{figure}
We sketch the calculation of this amplitude,
which closely follows the approach of Bern and Morgan~\cite{Bern:1995db}. The relevant component tree diagrams can be 
extracted from Fig.~\ref{Hgg}. The left-hand side of the diagram is the colour-ordered amplitude for the $qgg\bar{q}$
process with positive helicity gluons which is given by,
\beq \label{Twogluonamplitude}
G_2^{\tree}(a,1^+, 2^+,b) = \frac{\spb1.2}{\spa1.2} \frac{\bar{u}(p_a) \gamma_R (\slsh{\mu}+m) u(p_b) }{(s_{a1}-m^2)},\;\;\;\gamma_R=(1+\gamma_5)/2\, .
\eeq
The components of the $d$-dimensional momenta $p_a$ beyond four dimensions are denoted by $\mu$ and $s_{ai}^2=(p_a+p_i)^2$.
With the normalization defined by Eq.~(\ref{qHqb}) the right-hand side of the diagram in Fig.~\ref{Hgg} is given by
\beq \label{qHqb}
H_0^{\tree} = m \bar{u}(p_b) u(p_a)\; .
\eeq
Sewing Eqs.~(\ref{Twogluonamplitude}) and (\ref{qHqb}) together and summing over the polarizations of fermions $a$ and $b$ we get
in four dimensions,
\beq
 m^2\frac{\spb1.2}{\spa1.2} \frac{{\rm Tr}\{\gamma_R (\sp_b +m) (\sp_a +m)\}}{(s_{a1}-m^2)}
=m^2\frac{\spb1.2}{\spa1.2} \frac{(2 p_a\cdot p_b +2 m^2)}{(s_{a1}-m^2)}
=m^2\frac{\spb1.2}{\spa1.2} \frac{(4 m^2-M_h^2)}{(s_{a1}-m^2)} \, .
\eeq
Restoring the propagators which were put on shell and exploiting the linkage between the mass terms and $\mu$,
we obtain a result for the amplitude, evaluated on the two particle cut,
\beq
A_2(1^+,2^+;H)_{scut}= m^2\frac{\spb1.2}{\spa1.2} \frac{1}{i \pi^2} \int d^d l\frac{(4 (m^2+\mu^2)-M_h^2)}{(l^2-m^2)((l+p_1)^2-m^2)((l+p_{12})^2-m^2)} \, .
\eeq
The symbol $p_i$ denotes the four-momentum of the $i$th particle,
and we further define, $p_{ij}=p_i+p_j$, $p_{ijk}=p_i+p_j+p_k$, etc.
Adding in the other diagram $1 \leftrightarrow 2$, and evaluating the rational term from $\mu^2$ we obtain,
\begin{equation} 
A_2 (1_g^+,2_g^+;H)= 2 m^2 \frac{\spb1.2}{\spa1.2} \Big[(4m^2-M_h^2)  C_0(p_1,p_2;m)+2\Big] \; .
\end{equation} 
where $C_0$ is the scalar triangle integral, defined in
Eq.(\ref{Integral_defns}). Note that the essential feature leading to
the simple answer was the simplified form of the tree level inputs.
In the following section we present the tree-level 
building blocks for one-loop Higgs amplitudes with greater numbers of
gluons.

\section{Tree level ingredients and cut techniques}
\subsection{Born-level results for $Q\bar{Q}$+$n$ gluon amplitudes}
Multi-gluon tree amplitudes with a pair of massive fermions have been considered by a number 
of authors~\cite{Ferrario:2006np,Ozeren:2006ft,Huang:2012gs} using BCFW techniques and supersymmetric
relations to scalar amplitudes. However since these authors make specific choices of spinors for the
massive fermions they are not well suited for our purposes.  
All orders results for tree graphs with $n$ gluons have been given in a convenient form in ref.~\cite{Ochirov:2018uyq}.
In our notation the $n+2$-point amplitude for a quark-antiquark pair and $n$ positive-helicity gluons 
is given in four dimensions by,
\beq
   G_n(a \:\!\!,1^+\!,2^+\!,\dots,n^+\!,b)
    = m  \frac{ \bar{u}(a) \gamma_R u(b)  \; [1|\prod_{j=1}^{n-2}\!\big\{\,\sp_{a\dots j} \sp_{j+1}
                                        +(s_{a 1 \dots j}-m^2) \big\}|n] }
           { (s_{a1}\!-\!m^2)(s_{a12}\!-\!m^2)\dots(s_{a1\dots(n-1)}\!-\!m^2)\,
             \braket{12}\braket{23}\dots\braket{n\!-\!1|n} } .\!\!
\label{QQggnAP}
\eeq
The important features of the all-positive helicity gluon amplitude are that the amplitude vanishes for massless quarks 
and that spin structure of the dependence on the massive quark momenta enters through the 
combination $\bar{u}(a) \gamma_R u(b)$ for all $n$. 

For $n=2$ the product collapses to unity and we recover the four dimensional version of Eq.~(\ref{Twogluonamplitude})
\beq
   G_2(a,1^+,2^+,b)
    = m 
   \frac{\bar{u}(a) \gamma_R u(b)} { (s_{a1}-m^2)} \frac{\spb1.2}{\spa1.2 } \, .
 \label{n=2}
\eeq
The results for larger numbers of gluons are similarly compact. For example, for $n=3,4$ we obtain,
\beqn
   G_3(a,1^+,2^+,3^+,b)
    &=& m  \frac{\bar{u}(a) \gamma_R u(b)\; [1|\left(\sp_{a1} \sp_{2}+(s_{a1}-m^2) \right)|3] }
           {(s_{a1}-m^2)(s_{a12}-m^2)\;\spa1.2 \spa2.3}\, .
\label{n=3} \\
   G_4(a,1^+,2^+,3^+,4^+,b)
    &=& m  \frac{\bar{u}(a) \gamma_R u(b)\; [1|
\left(\sp_{a1} \sp_{2}+(s_{a1}-m^2)\right)
\left(\sp_{a12} \sp_{3}+(s_{a12}-m^2)\right)|4] }
           {(s_{a1}-m^2)(s_{a12}-m^2)(s_{a123}-m^2)\;\spa1.2 \spa2.3 \spa3.4}\, .
\label{n=4}
\eeqn
\subsection{Interference with Higgs amplitudes}
From Eq.~(\ref{QQggnAP}) the all-positive helicity gluon amplitude has the same kinematic structure for all $n$.
It is therefore useful to contract the Higgs production amplitudes with this structure,
\begin{equation} \label{spinorstructure}
M_0^\dagger = m \; \bar{u}(b) \gamma_R u(a)
\end{equation}
We will interfere this structure with the amplitude for
the production of a Higgs + 0,1 or 2 gluons.  Summing over the
polarizations of the massive quarks 
\beq
\sum u(p) \bar{u}(p)= \sp+m\, ,
\eeq
we obtain the following results
for the interference with zero, one and two gluon amplitudes, $H_0$, $H_1$ and $H_2$,
\begin{equation} \label{qHq}
M_0^\dagger \; H_0(a;H,b)=m^2 \Big(4 m^2 -M_h^2\Big) \, ,
\end{equation}
\begin{equation} \label{qgHq}
M_0^\dagger \; H_1(a,1_g^+;H,b)=m^2 \Big(4 m^2 -M_h^2\Big) 
    \frac{1}{\spa1.q }\Big\{ \frac{[1|a|q\rangle }{[1|a|1\rangle}-\frac{[1|b|q\rangle }{[1|b|1\rangle}\Big\}
=m^2 \Big(4 m^2 -M_h^2\Big) \frac{[1|ab|1]}{[1|a|1\rangle \, [1|b|1\rangle} \, ,
\end{equation}
(where $q$ is an arbitrary light-like vector),
\beqn \label{qggHq}
&&M_0^\dagger \; H_2(a,1_g^+,2_g^+;H,b)=2 m^2 \Big(4 m^2 -M_h^2\Big) \frac{1}{\spa1.2 }
           \Big\{
           \frac{[2|\slsh{b}(\slsh{b}-\slsh{a}) |1]} {[1|a|1\rangle \, [2|b|2\rangle} \nonumber \\
            &-&m^2 \spb2.1 \Big( \frac{1}{[2|b|2\rangle\, ((b-p_{12})^2-m^2)}+\frac{1}{[1|a|1\rangle \, [2|b|2\rangle }
         -\frac{1}{[1|a|1\rangle \, ((a+p_{12})^2-m^2)}\Big)\Big\} \, .
\eeqn
We take all momenta to be outgoing except for $b$.
We note that in four dimensions the interference of Eq.~(\ref{spinorstructure}) with the Higgs + gluon amplitudes, $H_n$,
is always proportional to $4 m^2-M_h^2$. 

\subsection{Higgs + 4 gluon amplitude: the coefficient of the scalar pentagon}

In four dimensions a scalar pentagon integral can be expressed as a sum of five
boxes~\cite{Melrose:1965kb,vanNeerven:1983vr}. (This reduction formula is described
in  appendix~\ref{fivetofour}). 
Consequently any attempt to identify the
coefficient of a pentagon integral is inherently a $d$-dimensional
calculation.  In $d$-dimensions we must introduce an extra parameter
$\mu$, that describes the magnitude of the loop momentum momentum
in the $(d-4)=-2 \epsilon$ space.  We now use unitarity to extract the
coefficient of the scalar pentagon integral for the diagram shown in
Fig.~\ref{Hgggg}.  We express the loop momentum $l$ as,
\beq \label{loopmomentum}
l^{\nu} = \alpha \, p_1^{\nu} + \beta \, p_2^{\nu} 
+\frac{\gamma}{2}\, \langle 1|\gamma^\nu | 2]+\frac{\delta}{2}\,\langle 2|\gamma^\nu | 1] + l_{\epsilon}^{\nu}.
\eeq
We denote the length of the component of $l$ beyond 4 dimensions, $l_{\epsilon}$, by $\mu$. 
Placing all five propagators on their mass shell we obtain the following five equations,
\beqn \label{fiveeqns}
 l^2-m^2=&0,             \;\;\;\to\;\;\;\;& -\gamma\delta \spa1.2\spb2.1 -m^2-\mu^2 =0\, ,~{{\rm determines}}~\mu^2\, , \nonumber \\
(l-p_1)^2-m^2=&0,        \;\;\;\to\;\;\;\;& \beta =0 \, , \nonumber \\
(l+p_2)^2-m^2=&0,        \;\;\;\to\;\;\;\;& \alpha =0 \, ,\nonumber \\
(l+p_2+p_3)^2-m^2=&0,    \;\;\;\to\;\;\;\;& \gamma \spa1.3 \spb3.2)+\delta \spa2.3 \spb3.1+s_{23}=0 \, , \nonumber \\
(l+p_2+p_3+p_4)^2-m^2=&0,\;\;\;\to\;\;\;\;& \gamma \spa1.4 \spb4.2+\delta \spa2.4 \spb4.1+s_{234}-s_{23}=0\, .
\eeqn
\begin{figure}[t]
\label{Hgggg}
\begin{center}
\includegraphics[width=4cm,angle=270]{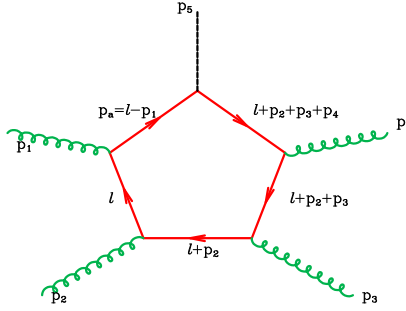}
\caption{\label{Hgggg} Feynman diagram to illustrate the calculation of coefficient the scalar pentagon integral.}
\end{center}
\end{figure}
However, because of the good ultraviolet properties of the pentagon integral, terms of order
less than $\mu^6$ will play no part in the limit $\epsilon \to 0$  and can be ignored.

The pentagon coefficient of Higgs plus four gluon amplitude in
Fig~(\ref{Hgggg}) can be calculated by putting all five propagators
on-shell and sewing together the $qgggg \bar{q}$ amplitude,
Eq.~(\ref{n=4}) and the projection of the Higgs production vertex,
Eq.~(\ref{qHq}). After imposing the mass-shell conditions, all dependence on the loop momentum drops out
and the result for the coefficient of $E_0(p_1,p_2,p_3,p_4;m)$ is,
\beq
m^2(4m^2-M_h^2)\frac{[1|\slsh{l}\,\slsh{p_2}\,(\slsh{l}+\slsh{p_2})\slsh{p_3}|4]}{\spa1.2\spa2.3\spa3.4} 
= -m^4(4m^2-M_h^2)\frac{\TrfourgR1.2.3.4}{\spa1.2\spa2.3\spa3.4\spa4.1} \, .
\eeq
To express this formula (and subsequent formula) we have introduced a
notation for the traces of gamma matrices (defined in detail in
Appendix~\ref{Traces}),
\beq
{\rm tr}_{+}\{1\, 2 \ldots n\} = {\rm tr} \{\gamma_R \,\sp_1\,\sp_2 \ldots \,\sp_{n}\}\, .
\eeq
The full result for the Higgs + 4 gluon amplitude is given in
Eq.~(\ref{Higgs4gluons}).

\subsection{Higgs + 5 gluon amplitude: the coefficient of one of the scalar pentagons}
We now use a similar method to identify the pentagon coefficient 
for the hexagon diagram shown in Fig.~\ref{Hggggg}.
\begin{figure}[t]
\label{Hggggg}
\begin{center}
\includegraphics[width=4.5cm,angle=270]{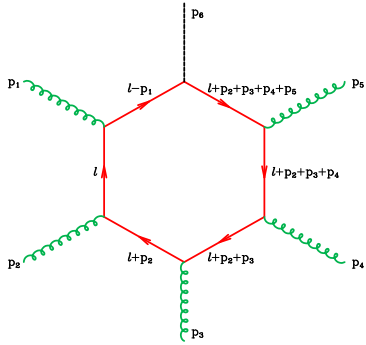} \\
\caption{\label{Hggggg} Feynman diagram to illustrate the calculation of the coefficient of the scalar integral $E_0(p_1,p_2,p_3,p_4;m)$}
\end{center}
\end{figure}
We parameterize the loop momentum as before, Eq.~(\ref{loopmomentum}) and
set the same condition on the five propagators, Eq~(\ref{fiveeqns}). The condition on the propagator $l^2-m^2$
serves to fix the length of the loop momentum in the extra dimension.
Solving the simultaneous equations for $\gamma$ and $\delta$ we have that,
\beqn \label{gammadeltasoln}
\gamma &=&  +\frac{1}{{\rm tr}_{5}(1,2,3,4)} \Big[\spa2.3 \spb3.1 (s_{234}-s_{23})-\spa2.4 \spb4.1 s_{23}\Big] \nonumber \\
\delta &=&  -\frac{1}{{\rm tr}_{5}(1,2,3,4)} \Big[\spa1.3 \spb3.2 (s_{234}-s_{23})-\spa1.4 \spb4.2 s_{23}\Big] 
\eeqn

With this solution for the $\gamma,\delta$ in hand we can evaluate the sixth denominator, 
$d_{6}=(l+p_2+p_3+p_4+p_5)^2-m^2$.
The result for $d_6$ on the cut of the first five denominators is
\beq
d_6 = -\frac{{\rm tr}_{5}(1,2,3,4,5,6)}{{\rm tr}_{5}(1,2,3,4)} \, .
\eeq
Not surprisingly, this is inverse of the coefficient which occurs in the reduction of 
a scalar hexagon integral to the scalar pentagon integral 
formed by the first five propagators, see Eq.~(\ref{6to5coefficients}). 
By evaluating any possible numerators factors for the value of $l$ determined by our $\gamma,\delta$ 
from Eq.~(\ref{gammadeltasoln}) 
we obtain the coefficient of this particular pentagon integral in our real physical amplitude.
Thus determining pentagon coefficients is even easier than box coefficients, because
we deal with a linear rather than a quadratic equation. The full result for the
Higgs + 5 gluon amplitude is given below in Eq.~(\ref{Higgs5gluons}).

\section{Results for Higgs + gluon amplitudes with all positive helicity gluons}
\subsection{$n=2$}
For the case $n=2$ we have the well known result\cite{Georgi:1977gs,Wilczek:1977zn} 
\begin{equation} 
A_2 (1_g^+,2_g^+;H)= 2 m^2 \frac{\spb1.2}{\spa1.2} \Big[(4m^2-M_h^2)  C_0(p_1,p_2;m)+2\Big] \, .
\end{equation} 
For $n=2$ the same helicity amplitudes are the only non-zero amplitudes.
We follow the normal notation for spinor products,\cite{Dixon:2013uaa} with 
$\langle ij\rangle [ji]=s_{ij}$ where $s_{ij}=p_{ij}^2=2 p_i \cdot p_j$ for the lightlike momenta $p_i$ and $p_j$.
The $C_0$ functions are the scalar triangle integrals, defined along with the box, pentagon and hexagon integrals,
$D_0,E_0$ and $F_0$ in Eq.~(\ref{Integral_defns}). 

\subsection{$n=3$}
For the case $n=3$ the results for all helicities are given in ref.~\cite{Ellis:1987xu}. The result for all 
positive helicity gluons is given by,
\begin{eqnarray} \label{Higgs3gluons}
A_3 (1_g^+,2_g^+,3_g^+;H)&=&  m^2 \Bigg[\Bigg\{   \frac{4 m^2-M_h^2}{\spa1.2\spa2.3\spa3.1} 
      \Big[-\frac{1}{2} s_{12} s_{23} 
              D_0(p_1,p_2,p_3;m) \nonumber \\
    &-&(s_{12}+s_{13})
           C_0(p_1,p_{23};m)\Big]-2 \frac{s_{12}+s_{13}}{\spa1.2\spa2.3\spa3.1}\Bigg\} \nonumber \\
 &+&\Bigg\{ 2~{\rm cyclic~permutations}\Bigg\}\Bigg]\, .
\end{eqnarray}
This result of ref.~\cite{Ellis:1987xu} has been confirmed in ref.~\cite{Baur:1989cm} 
where it is presented in a notation similar to the notation of the current paper. 
This result has been also obtained later by unitarity methods in ref.~\cite{Rozowsky:1997dm}.

\subsection{$n=4$}
Analytical results for the full one-loop amplitude for Higgs + 4 gluons
have been calculated for all helicities by the authors of
ref.~\cite{Neumann:2016dny} and are available in MCFM.
However simple analytic results have not been achieved.
For the case $n=4$ we find the simple expression,
\begin{eqnarray} \label{Higgs4gluons}
A_4 (1_g^+,2_g^+,3_g^+,4_g^+;H)&=&  m^2 \Bigg[\Bigg\{ \frac{4 m^2-M_h^2}{\spa1.2\spa2.3\spa3.4\spa4.1} 
 \Big[- \TrfourgR1.2.3.4  m^2 E_0(p_1,p_2,p_3,p_4;m)   \nonumber \\
&+& \frac{1}{2} ((s_{12}+s_{13})(s_{24}+s_{34})-s_{14}s_{23}) 
  D_0(p_1,p_{23},p_4;m)\nonumber \\
&+&\frac{1}{2} s_{12} s_{23} D_0(p_{1},p_{2},p_{3};m)\nonumber \\
&+& (s_{12}+s_{13}+s_{14}) C_0(p_1,p_{234};m)\Big] +2 \frac{s_{12}+s_{13}+s_{14}}{\spa1.2\spa2.3\spa3.4\spa4.1} 
\Bigg\}\nonumber \\
&+&\Bigg\{ 3~{\rm cyclic~permutations}\Bigg\}\Bigg] \, .
\end{eqnarray}
 
\subsection{$n=5$}
For the case $n=5$ we find,
\begin{eqnarray} \label{Higgs5gluons}
&&A_5 (1_g^+,2_g^+,3_g^+,4_g^+,5_g^+;H)=
m^2 \Bigg[\Bigg\{  \frac{(4 m^2-M_h^2)}{\spa1.2\spa2.3\spa3.4\spa4.5\spa5.1} \Big[ \sum_{i=1}^6 e_{(i)} E_{(i)} \nonumber \\
 &-&\frac{1}{2} s_{12}s_{23} D_0(p_1,p_2,p_3;m)
 -\frac{1}{2} \left[ (s_{12}+s_{13})(s_{24}+s_{34})-s_{14}s_{23}\right] D_0(p_1,p_{23},p_4;m) \nonumber \\
 &-&\frac{1}{2} \left[ (s_{12}+s_{13}+s_{14})(s_{25}+s_{35}+s_{45})-s_{15}(s_{23}+s_{24}+s_{34})\right] D_0(p_1,p_{234},p_5;m) \nonumber \\
 &-& \left(s_{12}+s_{13}+s_{14}+s_{15}\right) C_0(p_1,p_{2345};m)\Big] 
 -\frac{2 (s_{12}+s_{13}+s_{14}+s_{15})}{\spa1.2\spa2.3\spa3.4\spa4.5\spa5.1} \Bigg\} \nonumber \\
 &+&\Bigg\{ 4~{\rm cyclic~permutations}\Bigg\}\Bigg]\,,
\end{eqnarray}
where the coefficients of the scalar pentagon integrals are given by,
\beqn \label{H5g_coeff}
      e_{(1)}&=&m^2 \Big[\frac{1}{2} \TrfourgL2.3.4.5 + \frac{s_{23} s_{34} s_{45} (\TrfourgL2.6.5.1+s_{51}s_{12})}{\Trsixgfive1.2.3.4.5.6} \Big]\,,
      \nonumber \\ 
      e_{(2)}&=&-m^2 s_{45} s_{34} \frac{\TrsixgL5.1.2.3.(1+2).6} {\Trsixgfive1.2.3.4.5.6}\,, \nonumber \\
      e_{(3)}&=&-m^2  \frac{\TrfourgR5.4.{(2+3)}.1 \, \TrsixgL1.2.3.4.5.6}  {\Trsixgfive1.2.3.4.5.6}\,, \nonumber \\
      e_{(4)}&=&-m^2  \frac{\TrfourgR1.2.(3+4).5 \,   \TrsixgL5.4.3.2.1.6}{\Trsixgfive5.4.3.2.1.6}\,, \nonumber \\
      e_{(5)}&=&-m^2 s_{12}s_{23} \frac{\TrsixgL1.5.4.3.(4+5).6} {\Trsixgfive5.4.3.2.1.6}\,, \nonumber \\ 
      e_{(6)}&=&m^2 \Big[\frac{1}{2} \TrfourgL4.3.2.1+\frac{s_{12}s_{23} s_{34} (\TrfourgL4.6.1.5+s_{45}s_{51})} {\Trsixgfive5.4.3.2.1.6} \Big]\,,
\eeqn
and the pentagon integrals $E_{(i)}\equiv F_0^{(i)}$ correspond to the
scalar hexagon integrals with the $i$th propagator removed, see
Eq.~(\ref{Fidefn}).  Note the absence of boxes of the form
$D_0(p_1,p_{23},p_{45};m)$, $D_0(p_{12},p_{3},p_{45};m)$,
$D_0(p_{1},p_{2},p_{345};m)$ and $D_0(p_{1},p_{2},p_{34};m)$ apart
from those which would occur if the scalar pentagons in
Eq.~(\ref{Higgs5gluons}) were expressed as a sum of boxes.  The
momentum of the Higgs boson is denoted by $p_6$ such that
\beq \label{Momentumconservation}
\sum_{i=1}^6 p_i=0\, .
\eeq
Note that $\Trsixgfive5.4.3.2.1.6=-\Trsixgfive1.2.3.4.5.6$. This relationship is important to show
that the apparent singularity in $ e_{(1)}$ and  $e_{(6)}$ in the limit $p_6 \to 0$ cancels, because
in that limit $E_0(p_1,p_2,p_3,p_4)=E_0(p_2,p_3,p_4,p_5)$.

\section{Limits}
One of the benefits of an analytic formula is that we can investigate the behaviour of the amplitudes
in various limits. In this section we shall present the behaviour of the amplitude in the limit of 
vanishing Higgs boson momentum and for large top mass. 
The high energy limits of Higgs + 4 parton amplitudes
have been considered in ref.~\cite{DelDuca:2003ba}. 

\subsection{Soft Higgs limit}
The insertion of a soft Higgs boson is performed by the operating on the corresponding multi-gluon amplitude without a Higgs boson
with the operator, 
\beq
\frac{m}{v} \frac{d}{dm} \equiv \frac{1}{v} 2 m^2 \frac{d}{dm^2}.
\eeq
The colour sub-amplitude for scattering of four positive helicity gluons via a loop of quarks has been presented by
Bern and Morgan\cite{Bern:1995db},
\beq
A_{4}(1^+_g,2^+_g,3^+_g,4^+_g) =
  -2 {\spb1.2 \spb3.4 \over \spa1.2 \spa3.4 } \Big[ m^4 D_0(p_1,p_2,p_3;m)-\frac{1}{6}\Big] \, .
\eeq
In the limit in which $p_5 \to 0$ the result for the four gluon + 
Higgs amplitude, Eq.~(\ref{Higgs4gluons}) 
including cyclic symmetrization can be written as,
\beqn
&& A_4 (1_g^+,2_g^+,3_g^+,4_g^+;H) \to -4 m^4 \frac{\spb1.2\spa2.3\spb3.4\spa4.1}{\spa1.2\spa2.3\spa3.4\spa4.1} \nonumber \\
&\times& \Big[\frac{1}{2} (D_0(p_1,p_2,p_3;m)+D_0(p_2,p_3,p_4;m)+D_0(p_3,p_4,p_1;m)+D_0(p_4,p_1,p_2;m)) \nonumber \\
         &+&m^2 (E_0(p_1,p_2,p_3,p_4;m)+E_0(p_2,p_3,p_4,p_1;m)+E_0(p_3,p_4,p_1,p_2;m)+E_0(p_4,p_1,p_2,p_3;m))\Big]\nonumber \\
&=& -2 \frac{\spb1.2\spa2.3\spb3.4\spa4.1}{\spa1.2\spa2.3\spa3.4\spa4.1} \left[4 m^4 D_0(p_1,p_2,p_3;m) +2 m^6 \frac{d}{dm^2} D_0(p_1,p_2,p_3;m)\right] \nonumber \\
&=& -2 \frac{\spb1.2\spa2.3\spb3.4\spa4.1}{\spa1.2\spa2.3\spa3.4\spa4.1} 2 m^2 \frac{d}{dm^2}\left[m^4 D_0(p_1,p_2,p_3;m)\right]\,,
\eeqn
since in the limit $p_5 \to 0$ we have that 
$\spb1.2\spa2.3\spb3.4\spa4.1 = -s_{12}s_{23}$ and
\beqn
\frac{d}{dm^2} \, D_0(p_1,p_2,p_3;m)&=&
 E_0(p_1,p_2,p_3,p_4;m)+E_0(p_2,p_3,p_4,p_1;m)\nonumber \\
&+&E_0(p_3,p_4,p_1,p_2;m)+E_0(p_4,p_1,p_2,p_3;m) \, .
\eeqn
This demonstrates the expected form in the limit $p_5 \to 0$.
Similarly Eq.~(\ref{Higgs5gluons}) can be studied in the limit $p_6 \to 0$. 
In fact, we make use of the existence of this limit to help 
organise coefficients of scalar pentagons presented in Eq.~(\ref{H5g_coeff}).

\subsection{Large top mass limit}
In the large top mass limit we obtain the following results for the scalar integrals
\beqn
C_0(p_1,p_2;m)&=&-\frac{1}{2 m^2}-\frac{(p_1^2+p_2^2+p_{12}^2)}{24 m^4} +O\left(\frac{1}{m^6}\right) \, , \\
\label{D0limit}
D_0(p_1,p_2,p_3;m)&=&\frac{1}{6 m^4}+\frac{(s_{23}+s_{12}+p_1^2+p_2^2+p_3^2+p_{123}^2)}{60 m^6} +O\left(\frac{1}{m^8}\right) \, , \\
E_0(p_1,p_2,p_3,p_4;m)&=&-\frac{1}{12 m^6}+O\left(\frac{1}{m^8}\right) \, .
\eeqn
Using these expansions we obtain the expected form~\cite{Dawson:1991au, Dixon:2004za}
for the tree graphs in the effective theory.
\beqn
A_2 (1_g^+,2_g^+;H)&=& +\frac{2}{3}\frac{M_h^4}{\spa1.2\spa2.1}\, , \\
A_3 (1_g^+,2_g^+,3_g^+;H)&=& -\frac{2}{3}\frac{M_h^4}{\spa1.2\spa2.3\spa3.1}\, , \\
A_4 (1_g^+,2_g^+,3_g^+,4_g^+;H)&=&+\frac{2}{3} \frac{M_h^4}{\spa1.2\spa2.3\spa3.4\spa4.1}\, , \\
A_5 (1_g^+,2_g^+,3_g^+,4_g^+,5_g^+;H)&=&-\frac{2}{3} \frac{M_h^4}{\spa1.2\spa2.3\spa3.4\spa4.5\spa5.1} \, .
\eeqn

\section{Conclusions}
The results of the paper have shown that, having simple expressions for the component
tree graph amplitudes in hand, it is feasible to extract compact expressions for the Higgs boson
+ $n$-parton amplitudes for $n \leq 5$. The results with all gluon helicities taken to be the same,
display simple patterns. One is tempted to try and extend these results to even higher $n$, but in view 
of the limited phenomenological importance of higher $n$ we have not succumbed to this temptation.
The results for $n=4$ and $n=5$, after extension to all helicities, offer
the prospect of fast and stable numerical evaluation.

\section*{Acknowledgements}
We would like to acknowledge useful discussions with Simon Badger and Nigel Glover. RKE 
gratefully acknowledges the hospitality and partial support of the Mainz Institute for Theoretical Physics (MITP) 
during the completion of this work.
\appendix
\section{Integrals}
\label{Integrals}
We define the denominators of the integrals as follows
\begin{eqnarray}
d_0 &= &l^2-m^2+i\varepsilon \, ,\nonumber\\
d_1 &= & (l+p_1)^2-m^2+i\varepsilon = (l+q_1)^2-m^2+i\varepsilon  \, ,\nonumber\\
d_{12} &= & (l+p_1+p_2)^2-m^2+i\varepsilon = (l+q_2)^2-m^2+i\varepsilon \, , \nonumber\\
d_{123} &= & (l+p_1+p_2+p_3)^2-m^2+i\varepsilon = (l+q_3)^2-m^2+i\varepsilon \, , \nonumber\\
d_{1234} &= & (l+p_1+p_2+p_3+p_4)^2-m^2+i\varepsilon = (l+q_4)^2-m^2+i\varepsilon \, , \nonumber\\ 
d_{12345} &= & (l+p_1+p_2+p_3+p_4+p_5)^2-m^2+i\varepsilon= (l+q_5)^2-m^2+i\varepsilon \, .
\end{eqnarray}
The $p_i$ are the external momenta, whereas the $q_i$ are the off-set momenta in 
the propagators. In terms of these denominators the integrals are,
\begin{eqnarray}
\label{Integral_defns}
C_0(p_1,p_2;m) &=& \frac{1}{i \pi^{2}} \int d^4l \frac{1}{d_0 d_1 d_{12}} \, ,\nonumber \\
D_0(p_1,p_2,p_3;m) &=& \frac{1}{i \pi^{2}} \int d^4l \frac{1}{d_0 d_1 d_{12}d_{123}} \, ,\nonumber \\
E_0(p_1,p_2,p_3,p_4;m) &=& \frac{1}{i \pi^{2}} \int d^4l \frac{1}{d_0 d_1 d_{12} d_{123} d_{1234}} \, ,\nonumber \\
F_0(p_1,p_2,p_3,p_4,p_5;m) &=& \frac{1}{i \pi^{2}} \int d^4l \frac{1}{d_0 d_1 d_{12} d_{123} d_{1234} d_{12345}} \, .
\end{eqnarray}

\section{Definitions of $\gamma$-matrix traces}
In order to obtain compact expressions for the coefficients of the scalar integrals, we define 
the following traces of $\gamma$-matrices.
\label{Traces}
\beqn \label{tracenotation}
{\rm tr}_{5}\{1\, 2 \ldots n\} &=& {\rm tr} \{\gamma_5 \,\sp_1\,\sp_2 \ldots \,\sp_{n}\} \, ,\nonumber \\
{\rm tr}_{+}\{1\, 2 \ldots n\} &=& {\rm tr} \{\gamma_R \,\sp_1\,\sp_2 \ldots \,\sp_{n}\} \, , \nonumber \\
{\rm tr}_{-}\{1\, 2 \ldots n\} &=& {\rm tr} \{\gamma_L \,\sp_1\,\sp_2 \ldots \,\sp_{n}\} \, ,\nonumber \\
{\rm tr}_{5}\{1\, 2 \ldots n\} &\equiv & {\rm tr}_{+}\{1\, 2 \ldots n\} - {\rm tr}_{-}\{1\, 2 \ldots n\} \, ,
\eeqn
with $\gamma_{R/L}=(1 \pm \gamma_5)/2$. For the special case of lightlike vectors we have that
\beqn
{\rm tr}_{+}\{1\, 2 \,3 \ldots n\} &=&  \spb1.2\,\spa2.3 \, \spb3.4\ldots \left\langle n \, 1\right\rangle \, ,\nonumber \\
{\rm tr}_{-}\{1 \, 2 \, 3\ldots n\} &=& \spa1.2\,\spb2.3 \, \spa3.4 \ldots \left[n \, 1 \right]\, .
\eeqn
In the case of lightlike vectors, the traces with $\gamma_5$ can be written as differences of spinor strings,
\beqn
{\rm tr}(\gamma_5\,\sp_1\,\sp_2\,\sp_3\,\sp_4) &=& \big(\spb1.2\,\spa2.3\,\spb3.4\,\spa4.1 - \spa1.2\,\spb2.3\,\spa3.4\,\spb4.1\big)\, , \\
{\rm tr}(\gamma_5\,\sp_1\,\sp_2\,\sp_3\,\sp_4\,\sp_5\,\sp_6) &=& \big(
   \spb1.2\,\spa2.3\,\spb3.4\,\spa4.5 \,\spb5.6 \,\spa6.1
  -\spa1.2\,\spb2.3\,\spa3.4\,\spb4.5 \,\spa5.6 \,\spb6.1\big)\, .
\eeqn
In the case where external vectors are not light-like, (e.g. in our case the Higgs momentum $p_6$), the spinor expressions must be 
modified using momentum conservation, e.g. Eq.~(\ref{Momentumconservation}) for the five gluon case.

\section{Reduction of scalar pentagon integrals to boxes}
\label{fivetofour}
The reduction of the scalar pentagon integrals $E_0$ to a sum of the five boxes obtained by removing 
one propagator has been presented in ref.~\cite{vanNeerven:1983vr}. We present the result here for completeness.
\beq
E_{0}(w^2-4 \Delta_4 m^2)=
E^{(1)}\, [2 \Delta_4-w\cdot(v_1+v_2+v_3+v_4)]
+E_0^{(2)}\,v_1\cdot w
+E_0^{(3)}\,v_2\cdot w
+E_0^{(4)}\,v_3\cdot w
+E_0^{(5)}\,v_4\cdot w \, ,
\label{PentagonInt}
\eeq
where the vectors $v_i$ are expressed in terms of the totally antisymeetric tensor $\varepsilon$,
\beqn
v_1^\mu &=& \varepsilon^{\mu,q_2,q_3,q_4},\;\;
v_2^\mu = \varepsilon^{q_1,\mu,q_3,q_4},\;\;
v_3^\mu = \varepsilon^{q_1,q_2,\mu,q_4},\;\;
v_4^\mu = \varepsilon^{q_1,q_2,q_3,\mu},\nonumber \\
w^\mu &=& r_1 v_1^\mu+r_2 v_2^\mu+r_3 v_3^\mu+r_4 v_4^\mu \, ,
\eeqn
and the box integrals are,
\beqn \label{Eidefn}
E_0^{(1)}&=&D_0(p_2,p_3,p_4;m) \, ,\nonumber  \\
E_0^{(2)}&=&D_0(p_{12},p_3,p_4;m)\, ,\nonumber  \\
E_0^{(3)}&=&D_0(p_1,p_{23},p_4;m)\, ,\nonumber  \\
E_0^{(4)}&=&D_0(p_1,p_2,p_{34};m)\, ,\nonumber  \\
E_0^{(5)}&=&D_0(p_1,p_2,p_3,;m)\, ,
\eeqn
where $p_{ij}=p_i+p_j$. 
The $r_i$ are the residues when the dot products 
of the offset momenta and the loop momenta,  $q_i\cdot l$,
are expressed in terms of differences of propagators,
\beqn
q_1.l &=& \frac{1}{2} [d_1-d_0-r_1], \;\;\; q_2.l = \frac{1}{2} [d_{12}-d_0-r_2] \, ,\nonumber \\
q_3.l &=& \frac{1}{2} [d_{123}-d_0-r_3],\;\;\; q_4.l = \frac{1}{2} [d_{1234}-d_0-r_4] \, .
\eeqn
\section{Reduction of scalar hexagons integrals to pentagons}
\label{sixtofive}
The reduction of the scalar hexagon integrals $F_0$ to a sum of the six pentagons obtained by removing 
one propagator can be derived following the techniques of ref.~\cite{Melrose:1965kb,vanNeerven:1983vr}.
We denote by $F_0^{(i)}$ the six pentagon
integrals obtained by removing the $i$th propagator from the hexagon integral,
\beq
F_0(p_1,p_2,p_3,p_4,p_5;m) = \sum_{i=1}^{6}\; c_{12345}(i) F_0^{(i)} \, .
\eeq
Explicitly we have that,
\beqn \label{Fidefn}
F_0^{(1)}&\equiv&E_{(1)}=E_0(p_2,p_3,p_4,p_5;m) \, ,\nonumber  \\
F_0^{(2)}&\equiv&E_{(2)}=E_0(p_{12},p_3,p_4,p_5;m)\, ,\nonumber  \\
F_0^{(3)}&\equiv&E_{(3)}=E_0(p_1,p_{23},p_4,p_5;m)\, ,\nonumber  \\
F_0^{(4)}&\equiv&E_{(4)}=E_0(p_1,p_2,p_{34},p_5;m)\, ,\nonumber  \\
F_0^{(5)}&\equiv&E_{(5)}=E_0(p_1,p_2,p_3,p_{45};m)\, ,\nonumber  \\
F_0^{(6)}&\equiv&E_{(6)}=E_0(p_1,p_2,p_3,p_4;m)\, ,
\eeqn
where $p_{ij}=p_i+p_j$. 
Translating the results of ref.~\cite{vanNeerven:1983vr} to the notation of
Eq.~(\ref{tracenotation})
we find~(see also ref.~\cite{Binoth:2005ff}),
\beqn \label{6to5coefficients}
c_{12345}^{(1)}&=&+\Trfourgfive2.3.4.5/\Trsixgfive1.2.3.4.5.6 \, ,\nonumber \\
c_{12345}^{(2)}&=&-\Trfourgfive{(1+2)}.3.4.5/\Trsixgfive1.2.3.4.5.6 \, ,\nonumber \\
c_{12345}^{(3)}&=&+\Trfourgfive1.{(2+3)}.4.5/\Trsixgfive1.2.3.4.5.6 \, ,\nonumber \\
c_{12345}^{(4)}&=&-\Trfourgfive1.2.{(3+4)}.5/\Trsixgfive1.2.3.4.5.6 \, ,\nonumber \\
c_{12345}^{(5)}&=&+\Trfourgfive1.2.3.{(4+5)}/\Trsixgfive1.2.3.4.5.6 \, ,\nonumber \\
c_{12345}^{(6)}&=&-\Trfourgfive1.2.3.4/\Trsixgfive1.2.3.4.5.6 \, .
\eeqn
In this equation we have used an obvious extension of the notation of Eq.(\ref{tracenotation}),
\beq
\Trfourgfive{(1+2)}.3.4.5 \equiv {\rm tr}(\gamma_5\,(\sp_1+\sp_2)\,\sp_3\,\sp_4\,\sp_5)\, .
\eeq
Expressed in this form it is manifest that
\beq
 \sum_{i=1}^{6}\; c_{12345}^{(i)} =0 \, .
\eeq
\bibliography{poshel}
\bibliographystyle{JHEP}
\end{document}